\documentclass[a4paper,11pt]{article}
\usepackage{pos}
\usepackage{bbm}

\title{Colour--kinematics duality, double copy, and homotopy algebras}

\author[a]{Leron~Borsten}
\author[b]{Hyungrok~Kim}
\author[c]{Branislav~Jur{\v c}o}
\author*[d]{Tommaso~Macrelli}
\author[b]{Christian~Saemann}
\author[e]{Martin~Wolf}

\affiliation[a]{Department of Physics, Astronomy and Mathematics, University of Hertfordshire,
Hatfield, Hertfordshire, AL10 9AB, United Kingdom}
\affiliation[b]{Maxwell Institute for Mathematical Sciences,\\ Department of Mathematics, Heriot--Watt University, Edinburgh EH14 4AS, UK}
\affiliation[c]{Charles University Prague, Faculty of Mathematics and Physics,\\ Mathematical Institute, Prague 186 75, CZ}
\affiliation[d]{Institute for Theoretical Physics, ETH Zurich, 8093 Z{\"u}rich, Switzerland}
\affiliation[e]{Department of Mathematics, University of Surrey, Guildford GU2 7XH, UK}

\emailAdd{l.borsten@herts.ac.uk}
\emailAdd{hk55@hw.ac.uk}
\emailAdd{branislav.jurco@gmail.com}
\emailAdd{tmacrelli@ phys.ethz.ch}
\emailAdd{c.saemann@hw.ac.uk}
\emailAdd{m.wolf@surrey.ac.uk}

\abstract{Colour--kinematics duality is a remarkable property of Yang--Mills theory. Its validity implies a relation between gauge theory and gravity scattering amplitudes, known as double copy. Albeit fully established at the tree level, its extension to the loop level is conjectural. Lifting the on-shell, scattering amplitudes-based description to the level of action functionals, we argue that a theory that exhibits tree-level colour--kinematics duality can be reformulated in a way such that its loop integrands manifest a generalised form of colour--kinematics duality. Moreover, we show how the structures of higher homotopy theory naturally describe this off-shell reformulation of colour--kinematics duality.}

\FullConference{%
  41st International Conference on High Energy physics - ICHEP2022\\
  6-13 July, 2022\\
  Bologna, Italy
}

\RequirePackage{pgffor}

\newcommand{\makecommand}[3]{%
	\foreach \i in #3 {%
    	\expandafter\xdef\csname #1\i\endcsname{{\noexpand#2{\unexpanded\expandafter{\i}}}}%
  	}%
}

\newcommand{\latinalphabet}{A,a,B,b,C,c,d,D,E,e,F,f,G,g,H,h,I,i,J,j,K,k,L,l,M,m,N,n,O,o,P,p,Q,q,R,r,S,s,T,t,U,u,V,v,W,w,X,x,Y,y,Z,z}
\newcommand{\Latinalphabet}{A,B,C,D,E,F,G,H,I,J,K,L,M,N,O,P,Q,R,S,T,U,V,W,X,Y,Z}

\makecommand{I}{\mathbbm}{\latinalphabet}
\makecommand{bf}{\mathbf}{\latinalphabet}
\makecommand{bm}{\bm}{\latinalphabet}
\makecommand{ca}{\mathcal}{\Latinalphabet}
\makecommand{fr}{\mathfrak}{\latinalphabet}
\makecommand{rm}{\mathrm}{\latinalphabet}
\makecommand{sf}{\mathsf}{\latinalphabet}
\makecommand{sc}{\mathscr}{\latinalphabet}
\makecommand{tt}{\mathtt}{\latinalphabet}

\makecommand{bar}{\bar}{\latinalphabet}
\makecommand{dot}{\dot}{\latinalphabet}

\newcommand{\id}{\mathsf{id}}

\newcommand{\colfac}{\sfc}
\newcommand{\kinfac}{\sfn}

\let\oldblacksquare\blacksquare
\newcommand{\BBox}{{\textcolor{gray}\oldblacksquare}}
\newcommand{\BVbox}{BV^\BBox}

\usepackage{enumerate}

\begin{document}
\maketitle

\section{Introduction}
Yang--Mills theory and Einstein--Hilbert gravity are very different theories: while the former is a renormalisable theory of interacting massless spin 1 fields, invariant under the action of compact-Lie-group-valued functions, the latter is a non-renormalisable, diffeomorphism-invariant theory describing the dynamics of the space--time metric. If we consider the semi-classical perturbative description of said theories in terms of tree-level Feynman diagrams, the usual Yang--Mills action produces only cubic and quartic interaction vertices, while the Einstein--Hilbert action for gravity produces vertices of arbitrarily high degree.

It is surprising then to find that tree-level gravity scattering amplitudes can be obtained by ``squaring'' Yang--Mills amplitudes through a procedure known as double copy~\cite{Bern:2008qj,Bern:2010ue,Bern:2010yg}. This relation, which may be considered a reflection of Kawai--Lewellen--Tye relations that hold between closed and open string amplitudes, relies on a remarkable property of Yang--Mills theory, namely colour--kinematics duality. Consider an $n$-point, $L$-loop Yang--Mills amplitude $\scA_{n,L}$, parameterised as

        \begin{equation}\label{eq:YM_amplitudes_parameterization}
                \scA_{n,L}\ =\ (-\rmi)^{n-3+3L}g^{n-2+2L}\sum_{i\in \Gamma_{n,L}}\int\left(\prod^L_{l=1}\frac{\rmd^dp_l}{(2\pi)^d}\right)\frac{\colfac_i\kinfac_i}{S_id_i}~,
        \end{equation}
        where $g$ is the Yang--Mills coupling constant and $\Gamma_{n,L}$ is the set of cubic graphs\footnote{i.e.~graphs with vertices that all have degree three} with $n$ labelled external lines. The denominators $d_i$ are given by products of the Feynman--'t~Hooft propagators, i.e.~products of factors $\frac{1}{p^2_l}$ where $p_l$ is the momentum flowing through the internal line $l$. The $\colfac_i$ are colour factors, consisting of contractions of the gauge Lie algebra structure constants and the Killing form according to the structure of the graph $i\in \Gamma_{n,L}$. The kinematic factors $\kinfac_i$ are sums of Lorentz-invariant contractions of the Minkowski metric, external momenta and the polarisation vectors labelling the external scattering states.

We have some freedom in the choice of kinematic factors. Indeed, $\scA_{n,L}$ is invariant under any shift $\kinfac_i\to\kinfac_i+\Delta_i$ with
        \begin{equation}\label{eq:YM_amplitudes_generalized_gauge_tfm}
                0\ =\ \sum_{i\in \Gamma_{n,L}}\int\left(\prod^L_{l=1}\frac{\rmd^dp_l}{(2\pi)^d}\right)\frac{\Delta_i\kinfac_i}{S_id_i}~.
        \end{equation}
We call these shifts generalised gauge transformations. We say that the integrands of $\scA_{n,L}$ are colour--kinematics dual if there is a choice of kinematic factors that obey the same algebraic identities (e.g., Jacobi identity) of their correspondent colour factors. Colour--kinematics duality holds if every amplitude is colour--kinematics dual.

Several theories exhibit this property. Tree-level colour--kinematics duality has been proved for Yang--Mills theory~\cite{Bjerrum-Bohr:2009ulz,Stieberger:2009hq} and the non-linear sigma model (NLSM)~\cite{Chen:2013fya,Du:2016tbc,Carrasco:2016ldy}, while full off-shell colour--kinematics duality was shown for (the currents of) self-dual Yang--Mills theory~\cite{Monteiro:2011pc} and Chern--Simons theory~\cite{Ben-Shahar:2021zww}. Colour--kinematics duality for Yang--Mills theory implies that if we replace the colour factors $\colfac_i$ in~\eqref{eq:YM_amplitudes_parameterization} with a second copy of the kinematic factor $\kinfac_i$, we obtain a gravity scattering amplitude~\cite{Bern:2008qj,Bern:2010ue,Bern:2010yg}. The double copy prescription is a perturbative all-loop statement, and its loop-level validity relies on loop-level colour--kinematics duality, which has not been proved so far.

To overcome the difficulties in proving loop-level colour--kinematics duality, we bootstrap the scattering amplitude paradigm to an off-shell, action-based description of colour--kinematics duality and the double copy. In the following we propose a weaker, generalised off-shell notion of colour--kinematics duality, implied by standard colour--kinematics duality at the tree level. We then provide a natural characterisation of colour--kinematics duality in terms of homotopy algebraic structures. Finally, we introduce the notion of syngamy to generalise the double copy procedure, and we describe a homotopy algebraic interpretation of colour--kinematics duality. 

\section{Generalised colour--kinematics duality}
Loop integrands are unphysical objects, affected by field redefinitions. This motivates us to state that a theory described by an action functional $S$ satisfies generalised colour--kinematics duality if there exists an action \emph{semi-classically equivalent} to $S$ (that is, an action obtained from $S$ via field redefinitions and introducing auxiliary fields) such that the associated Feynman diagram expansion produces colour--kinematics dual loop integrands~\cite{Borsten:2021rmh}.

The main result of~\cite{Borsten:2021rmh} is that tree-level colour--kinematics duality implies a generalised form of colour--kinematics duality, in which colour--kinematics duality is realised up to potential counterterms. Starting from a colour--kinematics dual parameterisation of the tree-level scattering amplitudes, it is possible to produce a semi-classically equivalent action for the theory that manifests generalised colour--kinematics duality up to any desired perturbative order:

        \begin{enumerate}\itemsep-2pt
            \item Consider a tree-level colour--kinematics dual quantum field theory; choose a colour--kinema\-tics dual parameterisation of tree-level scattering amplitudes. We explicitly allow terms in $S$ that vanish once algebraic relations for the structure constants (e.g.~anti-symmetry and the Jacobi identity of the Lie algebra structure constants as well as symmetry and invariance of the metric) are taken into account. Set $S_{3}=S$, where $S_{n}$ denotes the action with off-shell, generalised colour--kinematics duality manifest up to $n$ points. Now proceed with the algorithm starting at  $n=4$. 
            \item\label{item:compareFeynmanDiagrams} The $n$-point tree Feynman diagrams produced by $S_{n-1}$ partition the $n$-point tree scattering amplitude into pieces corresponding to different pole structures produced by propagators corresponding to internal edges. Compare these partitions to the colour--kinematic-dual parameterisation of the $n$-point scattering amplitudes. The sum of the differences must vanish, as the tree-level scattering amplitudes must agree.
            \item Add the (vanishing) sum of the differences to the action $S_{n-1}$ as an $n$-point vertex, producing the action $S^{\text{on-shell}}_n$. The $m$-point tree Feynman diagrams produced by $S^{\text{on-shell}}_n$ then agree with those in the colour--kinematic-dual parameterisation of the $m$-point scattering amplitudes for all $m\leq n$. Observe that the action $S^{\text{on-shell}}_n$ may contain non-local terms of the form $A\frac{1}{\Box}B$.
            \item From $S^{\text{on-shell}}_n$, we compute off-shell amputated correlators. When restricted on shell, they manifest colour--kinematics duality: this implies that off-shell generalised colour--kinematics duality is violated by terms proportional to the equations of motion. These terms can be compensated by (possibly non-local) field redefinitions.
            \item If we have reached the maximal order that is of relevance for the scattering amplitudes we are interested in, halt. Otherwise, increment $n$, and go back to~\ref{item:compareFeynmanDiagrams}. 
        \end{enumerate}
Explicit computations for Yang--Mills theory and NLSM are detailed in~\cite{Borsten:2021rmh}.

\section{Double copy}
The action obtained at the end of the algorithm described in the last Section can be cast in a cubic form with the introduction of auxiliary fields
        \begin{subequations}\label{eq:left_theory}
            \begin{equation}\label{eq:left_cubic_action_decomposition}
                S=\tfrac12\sfg_{\alpha\beta}\,\bar\sfg_{\bar\alpha\bar\beta}\Phi^{\alpha\bar\alpha}\Box\Phi^{\beta\bar\beta}+\tfrac1{3!}\sff_{\alpha\beta\gamma}\,\bar\sff_{\bar\alpha\bar\beta\bar\gamma}\Phi^{\alpha\bar\alpha}\Phi^{\beta\bar\beta}\Phi^{\gamma\bar\gamma}~.
            \end{equation}
If we consider Yang--Mills theory or the NLSM, $\alpha,\beta,\gamma$ are colour or flavour indices, and $\bar\alpha,\bar\beta,\bar\gamma$ label particle species. The statement that generalised off-shell colour--kinematics duality is manifest is equivalent to the statement that $\sfg_{\alpha\beta},\sff_{\alpha\beta\gamma}$ and $\bar\sfg_{\bar\alpha\bar\beta},\bar\sff_{\bar\alpha\bar\beta\bar\gamma}$ satisfy the same algebraic relations. Motivated by the factorisation into left and right indices, we assume the BRST operator of the theory to be of the form
            \begin{equation}\label{eq:left_BRST_action}
                Q_{\text{BRST}}\Phi^{\alpha\bar\alpha}=\sfq^{\alpha}_{\beta}\delta^{\bar\alpha}_{\bar\beta}\Phi^{\beta\bar\beta}+\delta^{\alpha}_{\beta} \bar\sfq^{\bar\alpha}_{\bar\beta}\Phi^{\beta\bar\beta}+\tfrac12\sff_{\beta\gamma}{}^{\alpha}\,\bar\sfq^{\bar\alpha}_{\bar\beta\bar\gamma}\Phi^{\beta\bar\beta}\Phi^{\gamma\bar\gamma}+\tfrac12\sfq^{\alpha}_{\beta\gamma}\,\bar\sff_{\bar\beta\bar\gamma}{}^{\bar\alpha}\Phi^{\beta\bar\beta}\Phi^{\gamma\bar\gamma}+\cdots~.
            \end{equation}    
        \end{subequations}
        
       Consider a second theory that has manifest colour--kinematics duality (possibly a copy of the first one)
        \begin{subequations}\label{eq:right_theory}
            \begin{equation}\label{eq:right_cubic_action_decomposition}
                S=\tfrac12\sfg_{\sfa\sfb}\,\bar\sfg_{\bar\sfa\bar\sfa}\Phi^{\sfa\bar\sfa}\Box\Phi^{\sfb\bar\sfb}+\tfrac1{3!}\sff_{\sfa\sfb\sfc}\,\bar\sff_{\bar\sfa\bar\sfb\bar\sfc}\Phi^{\sfa\bar\sfa}\Phi^{\sfb\bar\sfb}\Phi^{\sfc\bar\sfc}
            \end{equation}
            with a BRST operator 
            \begin{equation}\label{eq:right_BRST}
                Q_{\text{BRST}}\Phi^{\sfa\bar\sfa}=\sfq^{\sfa}_{\sfb}\delta^{\bar\sfa}_{\bar\sfb}\Phi^{\sfb\bar\sfb}+\delta^{\sfa}_{\sfb}\bar\sfq^{\bar\sfa}_{\bar\sfb}\Phi^{\sfb\bar\sfb}+\tfrac12\sff_{\sfb\sfc}{}^{\sfa}\,\bar\sfq^{\bar\sfa}_{\bar\sfb\bar\sfc}\Phi^{\sfb\bar\sfb}\Phi^{\sfc\bar\sfc}+\tfrac12\sfq^{\sfa}_{\sfb \sfc}\,\bar\sff_{\bar\sfb\bar\sfc}{}^{\bar\sfa}\Phi^{\sfb\bar\sfb}\Phi^{\sfc\bar\sfc}+\cdots~.
            \end{equation} 
         \end{subequations} 
                 As detailed in~\cite{Borsten:2020zgj,Borsten:2021hua,Borsten:2021rmh}, we can combine the left and right components of all the fields and all the structure constants of the parent theories to obtain new theories. We refer to this generalisation of the double copy as \emph{syngamy}. We have four possible syngamies, each one with its own action and BRST operator:
        \begin{equation}\label{eq:gen_syngamies}
            \begin{aligned}
                \text{(i)}~~~&\Phi^{\alpha \sfa}~,~&&(\sfg_{\alpha\beta},\sfg_{\sfa\sfb},\sff_{\alpha\beta\gamma},\sff_{\sfa\sfb\sfc})~,~&&(\sfq^{\alpha}_{\beta},\sfq^{\sfa}_{\sfb},\ldots)~,
                \\
                \text{(ii)}~~~&\Phi^{\bar \alpha\sfa}~,~&&(\bar \sfg_{\bar\alpha\bar\beta},\sfg_{\sfa\sfb},\bar \sff_{\bar \alpha\bar \beta\bar \gamma},\sff_{\sfa\sfb\sfc})~,~&&(\bar\sfq^{\bar \alpha}_{\bar\beta},\sfq^{\sfa}_{\sfb},\ldots)~,
                \\
                \text{(iii)}~~~&\Phi^{\alpha \bar\sfa}~,~&&(\sfg_{\alpha\beta},\bar\sfg_{\bar\sfa\bar\sfb},\sff_{\alpha\beta\gamma},\bar\sff_{\bar\sfa\bar\sfb\bar\sfc})~,~&&(\sfq^{\alpha}_{\beta},\bar\sfq^{\bar\sfa}_{\bar\sfb},\ldots)~,
                \\
                \text{(iv)}~~~&\Phi^{\bar \alpha\bar\sfa}~,~&&(\bar \sfg_{\bar\alpha\bar\beta},\bar\sfg_{\bar\sfa\bar\sfb},\bar \sff_{\bar \alpha\bar \beta\bar \gamma},\bar\sff_{\bar\sfa\bar\sfb\bar\sfc})~,~&&(\bar\sfq^{\bar \alpha}_{\bar\beta},\bar\sfq^{\bar\sfa}_{\bar\sfb},\ldots)~.
            \end{aligned}
        \end{equation}

In the case of two Yang--Mills theories (identified with the left and right theories) with possibly different colour groups, these syngamies respectively correspond to:
        \begin{enumerate}[(i)]\itemsep-2pt
            \item A theory where the left kinematic part is combined with the right kinematic part. This is the syngamy that is usually called double copy. The resulting theory has the same field content as $\caN=0$ supergravity and is perturbatively quantum equivalent to this theory, cf.~\cite{Borsten:2021rmh}.
            \item A theory where the left colour part is combined with the right kinematic part. This syngamy is identical to the left parent theory. 
            \item A theory where the the left kinematic part is combined with the right colour part. This syngamy is identical to the right parent theory. 
            \item A theory where the left colour part is combined with the right colour part. This syngamy is sometimes called the zeroth copy, and the resulting theory is (quantum) equivalent to a theory of biadjoint scalars, cf.~\cite{Borsten:2021hua}.
        \end{enumerate}

\section{Homotopy algebraic description of colour--kinematics duality}
Every Lagrangian field theory can be described as a homotopy Maurer--Cartan theory associated to a cyclic $L_{\infty}$-algebra. In this final section, we employ this framework to characterise a theory whose scattering amplitudes admit a factorisation of the form~\eqref{eq:YM_amplitudes_parameterization}~\cite{Borsten:2022vtg}. Note that the algebraic structures that characterise colour--kinematics duality of Yang--Mills theory were also explored in the work of Reiterer~\cite{Reiterer:2019dys}. The field content of the theory is organised as a graded vector space $\frL=\oplus_{i\in\IZ}\frL_i$. The bilinear part of the action is described by a linear kinematic operator $\mu_1\colon\frL_\bullet\rightarrow \frL_{\bullet+1}$ of degree~$1$, and the cubic interactions are captured by graded-anti-symmetric bilinear maps $\mu_2\colon\frL\times \frL\rightarrow \frL$ of degree~$0$. Gauge symmetry and gauge invariance of the action imply that $(\frL,\mu_1,\mu_2)$ forms a differential graded Lie algebra, cf.~e.g.~\cite{Borsten:2021hua} for more details. The images of $\mu_1$ and $\mu_2$ appear in the action paired to the field content of the theory by a bilinear cyclic map of degree~$-3$. The kinematic operator $\mu_1$, restricted to a map from $\frL/\ker{\mu_1}$ to the image of $\mu_1$, is inverted by the propagator $P$: $ \mu_1P+P\mu_1+\id_\frL|_\text{on-shell}=\id_\frL$.
 Factorising $P=\frac{\sfh}{\Box}$ in terms of a linear operator $\sfh$ of degree $-1$, we can rewrite this equation as $\mu_1\sfh+\sfh\mu_1=\Box$. The colour-stripped theory is obtained after the factorisation $\frL=\frg\otimes \frB$, where $\frg$ is the colour Lie algebra and $\frB$ is a differential graded-commutative algebra with differential $\sfd$ and product $-\cdot-$~\cite{Borsten:2021hua}. Denoting the colour-stripped propagator also by $\frac{\sfh}{\Box}$, we have $\sfd\sfh+\sfh\sfd=\Box$.
 As we admit generic colour--kinematics dual theories in which the denominators $d_i$ appearing in~\eqref{eq:YM_amplitudes_parameterization} may be different than products of the squared momenta on internal legs, we allow a more general second order differential operator $\BBox$ with respect to $-\cdot-$. In all the relevant physical examples, $\sfh$ is also a second-order differential. We then have that
    \begin{equation}\label{eq:kinematic_bracket}
        [x,y]\coloneqq (-1)^{|x|}(\sfh(x\cdot y)-(\sfh x)\cdot y-(-1)^{|x|}x\cdot(\sfh y))
    \end{equation}
    is a Gerstenhaber bracket on $\frB$, i.e.~a Lie bracket of degree $-1$ mapping pairs of physical fields to physical fields, whose structure constants are precisely the kinematic structure constants. The kinematic Lie algebra is then $\frK\coloneqq(\frB[1],[-,-])$, where $\frB[1]$ indicates a shift of all fields by $-1$ needed to make $[-,-]$ an ordinary graded Lie bracket of degree~0. Observe that the kinematic Lie algebra usually discussed in the literature is obtained truncating $\frK$ to the elements of degree~0, the physical fields. Mathematically, $(\frB,-\cdot-,\sfh)$ forms a BV algebra in the sense of~\cite{Getzler:1994yd}, i.e.~a Gerstenhaber algebra $(\frB,-\cdot-,[-,-])$ in which the Gerstenhaber bracket $[-,-]$ is given by the derived bracket~\eqref{eq:kinematic_bracket}. The additional data enhance this BV algebra to what we call a $\BVbox$-algebra~\cite{Borsten:2022vtg} (see also~\cite{Akman:1995tm} for earlier generalisations of differential BV algebras).
\section*{Data Management}
No additional research data beyond the data presented and cited in this work are needed to validate the research findings in this work. For the purpose of open access, the authors have applied a Creative Commons Attribution (CC-BY) licence to any Author Accepted Manuscript version arising.


\bibliographystyle{plain}
\bibliography{bigone}

\begin{thebibliography}{10}

\bibitem{Akman:1995tm}
F{\"u}sun Akman.
\newblock {On some generalizations of {B}atalin--{V}ilkovsky algebras}.
\newblock {\em J. Pure Appl. Alg.}, 120:105--141, 1997.

\bibitem{Ben-Shahar:2021zww}
Maor Ben-Shahar and Henrik Johansson.
\newblock Off-shell color--kinematics duality for {C}hern--{S}imons.
\newblock {\em JHEP}, 2208:035, 2022.

\bibitem{Bern:2008qj}
Z.~Bern, J.~J.~M. Carrasco, and H.~Johansson.
\newblock New relations for gauge-theory amplitudes.
\newblock {\em Phys. Rev. D}, 78:085011, 2008.

\bibitem{Bern:2010ue}
Zvi Bern, John Joseph~M. Carrasco, and Henrik Johansson.
\newblock Perturbative quantum gravity as a double copy of gauge theory.
\newblock {\em Phys. Rev. Lett.}, 105:061602, 2010.

\bibitem{Bern:2010yg}
Zvi Bern, Tristan Dennen, Yu-tin Huang, and Michael Kiermaier.
\newblock Gravity as the square of gauge theory.
\newblock {\em Phys. Rev. D}, 82:065003, 2010.

\bibitem{Bjerrum-Bohr:2009ulz}
N.~E.~J. Bjerrum-Bohr, Poul~H. Damgaard, and Pierre Vanhove.
\newblock {Minimal Basis for Gauge Theory Amplitudes}.
\newblock {\em Phys. Rev. Lett.}, 103:161602, 2009.

\bibitem{Borsten:2022vtg}
Leron Borsten, Branislav Jurco, Hyungrok Kim, Tommaso Macrelli, Christian
  Saemann, and Martin Wolf.
\newblock {Kinematic Lie Algebras From Twistor Spaces}.
\newblock 11 2022.

\bibitem{Borsten:2020zgj}
Leron Borsten, Branislav Jur\v{c}o, Hyungrok Kim, Tommaso Macrelli, Christian
  Saemann, and Martin Wolf.
\newblock {BRST}--{L}agrangian double copy of {Y}ang--{M}ills theory.
\newblock {\em Phys. Rev. Lett.}, 126:191601, 2021.

\bibitem{Borsten:2021hua}
Leron Borsten, Branislav Jur\v{c}o, Hyungrok Kim, Tommaso Macrelli, Christian
  Saemann, and Martin Wolf.
\newblock {Double copy from homotopy algebras}.
\newblock {\em Fortsch. Phys.}, 69:2100075, 2021.

\bibitem{Borsten:2021rmh}
Leron Borsten, Branislav Jur\v{c}o, Hyungrok Kim, Tommaso Macrelli, Christian
  Saemann, and Martin Wolf.
\newblock Tree-level color--kinematics duality implies generalised loop-level
  color--kinematics duality.
\newblock 2021.

\bibitem{Carrasco:2016ldy}
John Joseph~M. Carrasco, Carlos~R. Mafra, and Oliver Schlotterer.
\newblock Abelian $z$-theory: {NLSM} amplitudes and $\alpha'$-corrections from
  the open string.
\newblock {\em JHEP}, 1706:093, 2017.

\bibitem{Chen:2013fya}
Gang Chen and Yi-Jian Du.
\newblock Amplitude relations in non-linear sigma model.
\newblock {\em JHEP}, 1401:061, 2014.

\bibitem{Du:2016tbc}
Yi-Jian Du and Chih-Hao Fu.
\newblock Explicit bcj numerators of nonlinear sigma model.
\newblock {\em JHEP}, 1609:174, 2016.

\bibitem{Getzler:1994yd}
E.~Getzler.
\newblock {Batalin--Vilkovisky algebras and two-dimensional topological field
  theories}.
\newblock {\em Commun. Math. Phys.}, 159:265--285, 1994.

\bibitem{Monteiro:2011pc}
Ricardo Monteiro and Donal O'Connell.
\newblock The kinematic algebra from the self-dual sector.
\newblock {\em JHEP}, 1107:007, 2011.

\bibitem{Reiterer:2019dys}
Michael Reiterer.
\newblock A homotopy {BV} algebra for {Y}ang--{M}ills and color--kinematics.
\newblock 2019.

\bibitem{Stieberger:2009hq}
S.~Stieberger.
\newblock {Open \& closed vs.~pure open string disk amplitudes}.
\newblock 2009.

\end{thebibliography}



\end{document}